\documentclass[arguments,twocolumn]{aastex631}

\usepackage{CJK}
\usepackage{graphicx}

\begin{document}

\begin{CJK*}{UTF8}{gbsn}

\title{The destiny of open cluster NGC 6530: past and future}

\correspondingauthor{Heng Yu}
\email{yuheng@bnu.edu.cn}

\author{Delong Jia(贾德龙)}
\affiliation{School of Physics and Astronomy, Beijing Normal University, Beijing, 100875, China}

\author[0000-0001-8051-1465]{Heng Yu(余恒)}
\affiliation{School of Physics and Astronomy, Beijing Normal University, Beijing, 100875, China}

\author[0000-0001-8611-2465]{Zhengyi Shao(邵正义)}
\affiliation{Shanghai Astronomical Observatory, Chinese Academy of Sciences, Shanghai, 200030, China}
\affiliation{Key Lab for Astrophysics, 100 Guilin Road, Shanghai, 200234, People’s Republic of China}

\author[0000-0002-0880-3380]{Lu Li(李璐)}
\affiliation{Shanghai Astronomical Observatory, Chinese Academy of Sciences, Shanghai, 200030, China}

\begin{abstract}
Studying the structures of open clusters is crucial for understanding  stellar evolution and galactic dynamics. Based on Gaia DR3 data, we apply the hierarchical clustering algorithm to the young open cluster NGC 6530 and group its members into five substructures. By linear tracing with the kinematic information of their members, we find that sub 1 is the core of the cluster. It is expanding slowly. Sub 2 consists of less-bound members, which began escaping from the core about 0.78 Myr ago. Sub 3 is associated with a young star forming region. It will merge with the core after 0.72 Myr. Sub 4, as an outskirts group, is also moving toward the core but will not end up falling in. While Sub 5 is composed of less-bound members with field contamination. 
This work reveals the complex internal structure and evolutionary trends of the cluster NGC 6530.  It also shows the potential of the hierarchical clustering algorithm in star cluster structure analysis.

\end{abstract}

\keywords{Open star clusters (1160) --- Single-linkage hierarchical clustering (1939) --- Stellar kinematics(1608) --- Astronomy data analysis (1858) }

\section{Introduction} \label{sec:intro}
Open star clusters typically consist of hundreds of stars that are loosely distributed in the sky\citep{2003ARA&A..41...57L}. As relatively young structures, they span an age range from a few million to several billion years\citep{2003AJ....126.1916P}. These stars originate from the same parent molecular cloud, retaining a physical connection with one another\citep{2008MNRAS.389..869E}. 
Open star clusters provide a natural laboratory for studying the processes of star formation and evolution. Their properties, such as ages and chemical compositions, serve as chronometers and tracers of the history of their surrounding environment.
Their distribution and kinematics can shed light on the nature of dark matter and the structure of the Milky Way \citep{2019ARA&A..57..227K}.

The initial step in studying star clusters is to determine their members.
There are many methods for identifying star cluster members. 
\cite{1958AJ.....63..387V} and \cite{1971A&A....14..226S} proposed the maximum-likelihood method based on the Gaussian distribution of cluster and field star proper motions. It is widely adopted after several revisions and improvements \citep{1990A&A...237...54Z, 1995AJ....109..672K, 2004A&A...416..125D, 2010A&A...516A...3K}. 
\cite{2014A&A...563A..45S} used a multidimensional Gaussian mixture model to select members of the Pleiades cluster.
\cite{2014ChA&A..38..257G} applied the DBSCAN algorithm to identify members of open clusters.
\cite{2014A&A...561A..57K} developed an unsupervised photometric membership assignment in
stellar clusters (UPMASK) method, which uses only photometry and positions to select cluster members. 
\cite{2019ApJS..245...32L} established a star cluster catalog using the friends-of-friends algorithm with Gaia DR2 data. 
Recently, the Hierarchical Density-Based Spatial Clustering of Applications with Noise (HDBSCAN) algorithm has also been used to identify cluster members \citep{2017JOSS....2..205M}, and \cite{2023A&A...673A.114H} conducted a systematic analysis of star clusters in the Milky Way using  Gaia DR3 data with this method.

High-quality stellar data from Gaia have advanced the study of clusters to a new stage.
Gaia is a space mission developed by the European Space Agency (ESA) to construct a precise three-dimensional map of our Milky Way galaxy. It measures positions, proper motions, and parallaxes of stars with high precision\citep{2023A&A...674A...1G}.

With the precise kinematic information of stars provided by Gaia, there have been many detailed studies on open clusters. 
\cite{2023A&A...671A.106M} studied the kinematic differences among different subpopulations within the cluster. \cite{2023MNRAS.526.1057R} used blue straggler stars' space distribution to analyze the kinematic ages of star clusters. %
\cite{2023MNRAS.523.5306F} utilized the consistency of proper motion between different groups to search for Collinder 121 and its surrounding structures. 
\cite{2023A&A...670A.128A} adopted unbound members of star clusters to determine the time of gas expulsion from the cluster. \cite{2023arXiv231002441V} performed traceback computations to identify stars that have recently escaped from clusters.

To explore the inner structure of clusters, \citet{YuShao-5} introduced the hierarchical clustering method, which utilizes the positions and proper motions of stars.
This method not only identifies cluster members but also reveals substructures and their hierarchical relationships.
It sheds light on the study of the cluster's internal structure and outskirts members.

In this study, we select NGC 6530 for analysis. NGC 6530 is a young cluster located within the Lagoon Nebula, with a distance of $1336^{+76}_{-68}$pc \citep{2019ApJ...870...32K}.
\cite{2019A&A...623A.159P} employed isochronal age lines fitted to color-magnitude diagrams (CMDs), indicating ages between 0.30 and 1.58 million yr. 
\cite{2019A&A...623A..25D} derived an age range between 0.5 and 5 million yr by combining isochronal age lines with spatial distribution.

In the sky plane, NGC 6530 exhibits an expanding core while surrounding structures are collapsing \citep{2019MNRAS.489.2694W}. In the western part of the cluster, there is the Hourglass Nebula \citep{2019MNRAS.486.2477W}, which contains numerous young stars and an exceptionally compact $\mathrm{H_{II}}$ region \citep{2008hsf2.book..533T}. 
There is another isolated group of young stars further west, about 24\arcmin\  away, which has been recognized as a member by \citet{2019A&A...623A..25D} and \cite{2023A&A...673A.114H}. Its relationship to the cluster is still unclear.

This complex structure of the cluster and its active environment make it an attractive object to explore. 
The structure of this paper is as follows. In Section \ref{sec:method}, we introduce the hierarchical clustering method for star cluster analysis. In Section \ref{sec:data}, we describe the data and present the results of member selection and substructure identification. 
In Section \ref{sec:trace}, we trace the evolution of substructures using kinematic information of their members.
In Section \ref{sec:diss}, we compare our kinematic ages with the age limitations obtained with isochrones on the CMD. 
Finally, conclusions are given in Section \ref{sec:conclusion}.

\section{Method} \label{sec:method}

In this section, we describe the hierarchical clustering method, which has been applied in prior research on the Perseus Double Cluster \citep{YuShao-5}.

This method arranges all stars in the field of view into a binary tree (dendrogram) based on the projected binding energy between stars. Stars with tighter binding appear deeper in the tree structure. 
By cutting the binary tree at suitable thresholds, we can extract kinematic structures from associated tree branches.

The projected binding energy between any two stars can be expressed as the sum of their gravitational potential energy and kinetic energy:

\begin{equation}
    E_{ij}=-G\frac{m_{i}m_{j}}{r\theta_{ij}}p+\frac{1}{2}\frac{m_{i}m_{j}}{m_{i}+m_{j}}\frac{{\triangle\mu_{x}}^2+{\triangle\mu_{y}}^2}{2}r^{2}
\end{equation}
 
\noindent In this equation, \textit{r} represents the distance to the star cluster, which is 1336pc.  $\theta_{ij}$ is the angle between two stars \textit{i} and \textit{j}. ${\triangle\mu_{x}}$ and ${\triangle\mu_{y}}$ are differences in the two orthogonal proper-motion components, where ${\triangle\mu_{x}=\triangle\mu_{RA}\cos(decl)}$ and ${\triangle\mu_{y}=\triangle\mu_{decl}}$. $m_{i}$ and $m_{j}$ represent the mass of the stars. 
Because the determination of $m_{i}$ is easily affected by the uncertainties in distance, luminosity, and other interfering factors associated with the star cluster, we set $m_{i}=m_{j}=1 M_{\odot}$.  The parameter \textit{p} is introduced to control the gravitational potential energy and kinetic energy within a reasonable range and will be determined according to the data (described in Section \ref{ssub}).

Then, we can construct the binary tree based on the projected binding energy between stars.
We identify the main branch of the tree by selecting the node with the largest number of leaves at each level of the tree. By traversing along this main branch and calculating the velocity dispersion associated with each node, we generate a velocity dispersion profile for the gravitational potential field. Once the branch enters the interior of the cluster, its velocity dispersion does not easily change with the decrease in members. A plateau will thus appear in the velocity dispersion profile. 
We refer to this plateau as the "$\sigma$ plateau," as in galaxy cluster studies\citep{1999MNRAS.309..610D, 2011MNRAS.412..800S, 2015ApJ...810...37Y}. 

To locate the plateau automatically, we fit the histogram of the velocity dispersion with a Gaussian mixture model.
We iterate over the number of Gaussian components in the range of 1-10 and take the value with the smallest Bayesian Information Criterion value as the optimal number.
When one cluster dominates the field, the Gaussian component with the largest weight ($w_0$) corresponds with the $\sigma$ plateau.
However, with different $p$, the position and length of the $\sigma$ plateau might change. To find a proper $p$ value for the dataset, we test different $p$ values and compare the corresponding $w_0$ values. When the $w_0$ value is maximized, the cluster is most clearly characterized. The corresponding $p$ value is adopted as the optimized value for the dataset \citep[see][for more details]{YuShao-5}.

\section{Data and Analysis} \label{sec:data}

\subsection{Data preparation}

Our data are selected from Gaia DR3 within a rectangular region centered at coordinates R.A.=$271.09^\circ$, decl.=$-24.33^\circ$, with a field of $1.2^\circ$ by $0.7^\circ$. We filter stars with a G-band magnitude brighter than 18 mag and exclude stars without proper motion and parallax information, resulting in a total of 20,819 stars as our initial data set.

When checking the initial data set, we find that there is a dominant structure in the field of view with a parallax of 0.4 $\rm{mas}$ (2500 $\rm{pc}$). This structure corresponds to the Scutum-Centaurus Arm of the Milky Way \citep{2021MNRAS.500.3064S}. To minimize the interference from these background stars, we further filter the initial data set with the NGC 6530 parallax range of $0.758 \pm 0.218$ $\rm{mas}$ (1024 -- 1851pc),  which is estimated using the method based on the initial data set (the information is described in the Appendix \ref{sec:appendixa}). In the end, we obtain 5945 stars for further analysis.

Out of the six kinematic parameters, four of them are applicable: two-dimensional celestial coordinates and two-dimensional proper motions. The average uncertainty in the parallax of these stars is 0.09 $\rm{mas}$, corresponding to a distance uncertainty of about 150 $\rm{pc}$ at the distance of the cluster. This value is much larger than the typical size of a cluster and thus does not provide an effective constraint. We do not take into account the radial velocity (the velocity of the star relative to the observer) because there are only 88 stars with radial velocities.

\subsection{Clustering}
\label{ssub}
\begin{figure}
\centering
\includegraphics[width=1\linewidth]{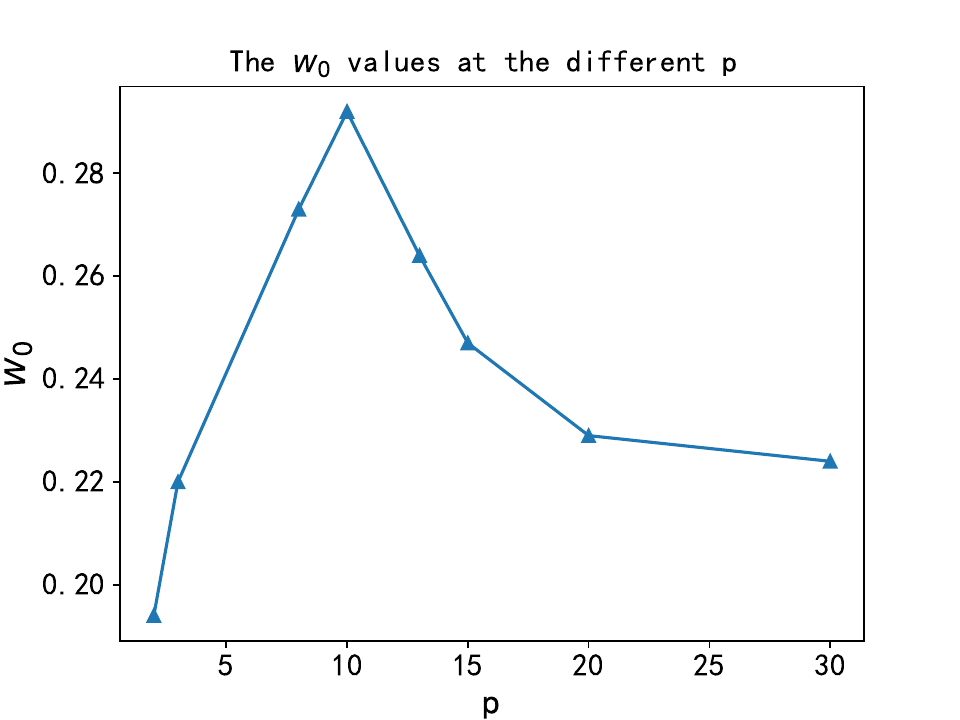}
\caption{\label{fig:w-p}The $w_0$-$p$ plot. $w_0$ represents the proportion of the largest plateau when the algorithm performs Gaussian fitting on different plateaus.}
\end{figure}

The first step of the hierarchical clustering algorithm is to determine the parameter $p$. We collect the weight values ($w_0$) of the $\sigma$-plateau with different $p$.  
The plot depicting the variation of $w_0$ with respect to the $p$ parameter is shown in Figure \ref{fig:w-p}. The figure reveals that the maximum $w_0$ value occurs when the $p$ parameter is set to 10 \citep[same as][]{YuShao-5}. Therefore, we set $p=10$ for further analysis.

\begin{figure}
\centering
\includegraphics[width=1\linewidth]{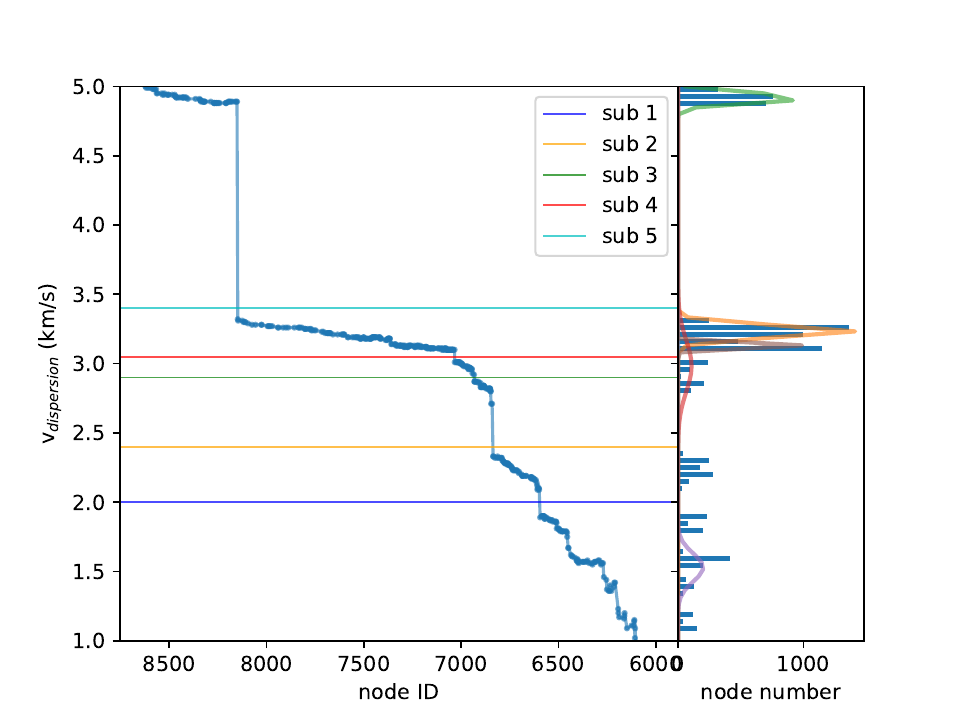}
\caption{\label{fig:vhst}{\it Left:} velocity dispersion profile, with leaf nodes on the x-axis and velocity dispersion values on the y-axis. 
The five colored lines, from low to high, represent 2.0, 2.4, 2.9, 3.05, and 3.4, respectively. 
{\it Right:} histogram of velocity dispersion, with solid colored lines representing multiple Gaussian components.}
\end{figure}

Our algorithm automatically recognizes the main velocity dispersion plateau around 3.2 $\rm{km~s^{-1}}$. At the beginning of the plateau, a threshold of 3.40 $\rm{km~s^{-1}}$   (the cyan line shown in Figure \ref{fig:vhst}) defines 895 members of the cluster.

There have been several other studies providing the membership of this cluster.
\cite{2019MNRAS.486.2477W} utilized the gravity index $\gamma$, the equivalent width if Li, and the full width at zero intensity of the $H_{\alpha}$ line to select members of NGC 6530, 
identifying 538 stars brighter than 18 mag in the $G$ band. \cite{2019A&A...623A..25D} used X-ray data, $H_{\alpha}$ emission, near-IR UV excesses, and Gaia DR2 astrometry to select 920 members of NGC 6530. 
\cite{2023A&A...673A.114H} selected 560 members of NGC 6530 using the HDBSCAN algorithm. 

We compare the properties of our members with those from previous studies as shown in Table  \ref{tab:compare}. 
For the members of \cite{2019A&A...623A..25D} and \cite{2019MNRAS.486.2477W}, we cross-match based on spatial coordinates using the 1\arcsec\ as the criterion. For members identified by \cite{2023A&A...673A.114H}, which are also from the Gaia DR3 data, our cross-match is based on the Gaia ID. The number of overlapping members is shown as the positive-positive (PP) value. Our algorithm covers most of their members. The minimum overlap ratio (the PP value over the catalog size) is 70\%, and the maximum overlap ratio is 93\%. 
Our members include more outskirts stars and inevitably introduce some contamination. Consequently, our velocity dispersion is larger than others.
The proper motions of our members are consistent with other studies in the decl. direction, but slightly larger in the R.A. direction.

\begin{table}
  \centering
  \caption{We use $n$ to represent the number of members selected by the reference method, and use PP (positive-positive) to represent the number of members jointly selected by our algorithm and the reference. Additionally, we provide the number of members with a G-band magnitude brighter than 18 and partial physical properties mentioned in the literature.}
  \label{tab:compare}
  {
  \begin{tabular}{|c|c|c|c|cc|}

    \hline
    Work & $n$ & PP & $v_{dis}$ & ${\mu}_{RA}$ & ${\mu}_{DEC}$ \\
         &      &         & ($\rm{km~s^{-1}}$ ) & ($\rm{mas~yr^{-1}}$ )   & ($\rm{mas~yr^{-1}}$ ) \\
    \hline
    W19 & 538  & 406 & 2.75 & 1.21 & -2.00 \\
    \hline
    D19 & 920  & 645 & 2.66  &  1.27 &  -2.00 \\
    \hline
    \protect{H23} & 560  &525 & 2.62 & 1.33 & -2.06 \\
    \hline
    Our & 895  & -- &  3.40 &  $1.41 \pm 0.43$ & $-2.03 \pm 0.28$ \\
    \hline
  \end{tabular}
  }
\end{table}

\subsection{Substructures}

In the velocity dispersion profile, there are jumps caused by the confluence of different structures. By cutting the tree around these jumps, we can extract substructures of the cluster.
There are four main jumps below 3.4 $\rm{km~s^{-1}}$ in the velocity dispersion profile of the main branch (Fig. \ref{fig:vhst}), which are located at about 2.0, 2.4, 2.9, and 3.05, respectively. There are also jumps below 2 $\rm{km~s^{-1}}$, but they are all related to the core of the cluster and differ very little in their physical properties. We will not discuss them further here.

Taking these four jumps (represented by four solid lines except cyan in Figure \ref{fig:vhst}) as cutting thresholds, we separate the cluster into five substructures:  sub 1 is a group of stars with a velocity dispersion less than 2.0,  sub 2 is members with a velocity dispersion between 2.0 and 2.4,  sub 3 is between 2.4 and 2.9, sub 4 is between 2.9 and 3.05, and sub 5 is between 3.05 and 3.4. 
Our substructures sometimes include a few members whose proper motions are close but whose spatial coordinates are significantly different. To improve the purity of substructures, we exclude members outside 3 times the half-mass radius (the radius containing half of the cluster members) in sub 1，2，3 and 4. Excluded substructure members, as members of the cluster, are assigned to sub 5.
\footnote{ The catalog of the members of NGC 6530 and its substructures derived with our procedure is publicly available at \href{https://paperdata.china-vo.org/yuheng/paper/NGC_6530_subid.txt}{https://paperdata.china-vo.org/yuheng/paper/NGC\_6530\_subid.txt}. }

\begin{figure*}
\centering
\includegraphics[width=1\linewidth]{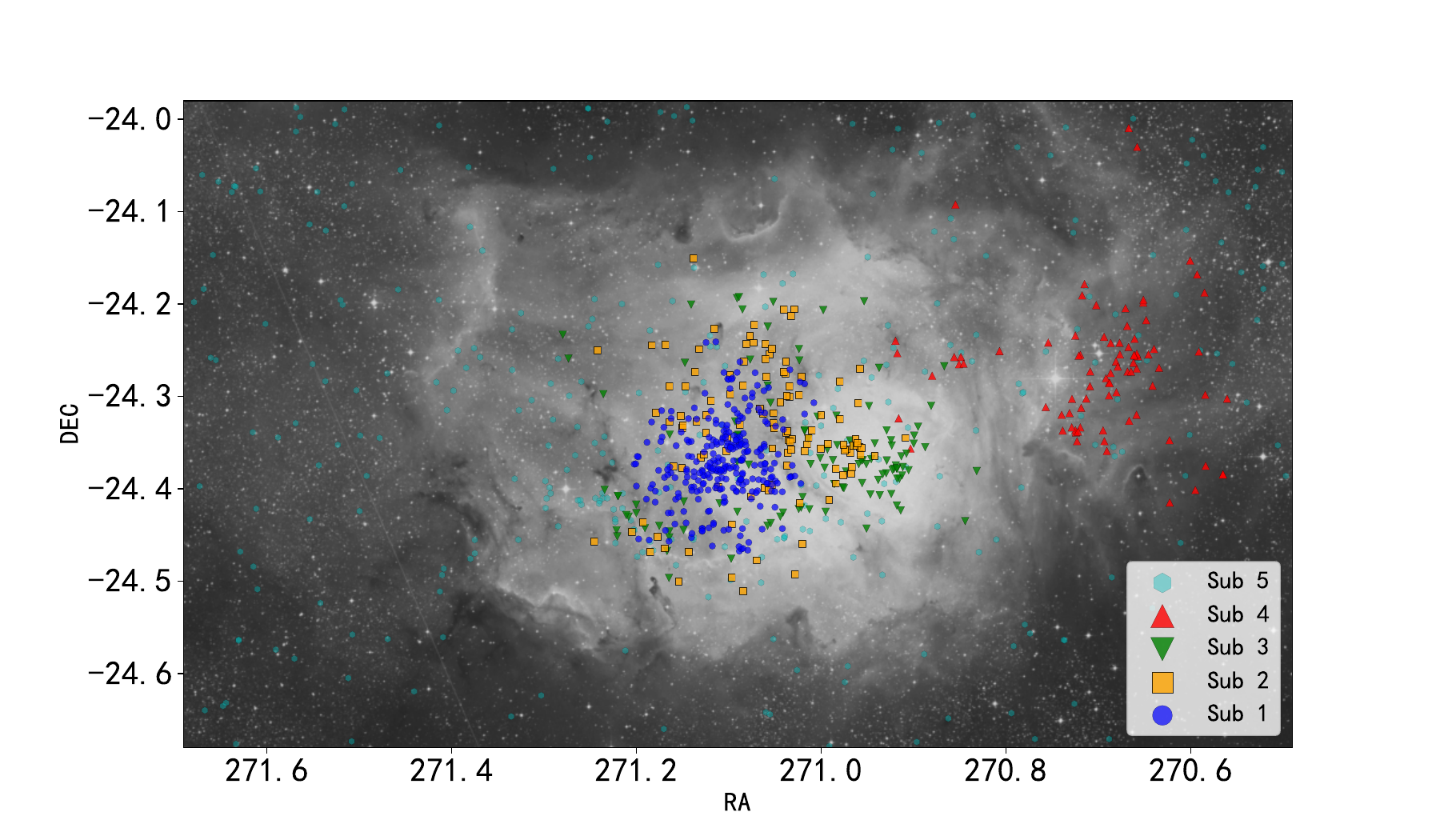}
\caption{\label{fig:space} The spatial distribution of five substructures of NGC 6530.
Sub 1 is represented by blue circles, sub 2 by orange squares, sub 3 by green inverted triangles, sub 4 by red triangles, and sub 5 by cyan hexagons. The background image is from the infrared band of the Digitized Sky Survey II (DSS2-infrared).}
\end{figure*}

The spatial distribution of these five substructures is shown in Figure \ref{fig:space}. Their basic physical properties are listed in Table \ref{tab:subs}.
Sub 1-2 is the combination of sub 1 and sub 2. They are both located in the core region of the cluster.
Considering our algorithm accepts many less-bound members in the outskirts of the cluster (mainly in sub 5), we also analyze the properties of member stars without sub 5.
Sub 1-4 contains 584 member stars below the $\sigma$ plateau (threshold 3.05 $\rm{km~s^{-1}}$). They are the most solid members of the cluster. 
 A total of 58\% of these members overlap with members of W19, 60 \% with those of D19 and 77 \% with those of H23.
The proper motions and velocity dispersions of sub1-4 are more consistent with the results of previous studies listed in Table \ref{tab:compare}.

\begin{table}[htpb]
    \centering
    \caption{Basic properties of substructures}
    \label{tab:subs}
    \begin{tabular}{|c|c|c|cc|}
        \hline
        & n  & $v_{dis}$& ${\mu}_{RA}$ & ${\mu}_{DEC}$  \\
        & & ($\rm{km~s^{-1}}$)& ($\rm{mas~yr^{-1}}$) & ($\rm{mas~yr^{-1}}$)   \\
        \hline
        Sub 1 & 275 & 1.89& $1.28\pm0.22$ & $-2.09\pm0.20$   \\
        Sub 2 & 114 & 2.90& $1.07\pm0.38$ & $-2.02\pm0.24$    \\
        Sub 3 & 111 & 3.32& $1.68\pm0.46$ & $-2.15\pm0.23$    \\
        Sub 4 & 84 & 1.76& $1.72\pm0.22$ & $-1.84\pm0.16$    \\
        Sub 5 & 311 & 3.87& $1.46\pm0.49$ & $-1.99\pm0.35$    \\
        \hline
        Sub 1-2 & 389 & 2.33& $1.22\pm0.29$ & $-2.07\pm0.21$    \\
        Sub 1-4 & 584 & 2.94& $1.38\pm0.39$ & $-2.05\pm0.23$    \\
        \hline
    \end{tabular}
\end{table}

The spatial distribution of the substructures is shown in Figure \ref{fig:space}.
Sub 1 lies in the center of the cluster, showing a compact and symmetrical morphology. 
Sub 2 is located on the outskirts of sub 1, being more diffuse and relatively concentrated in the northwest. 
The members of sub 3 are concentrated 0.2 degrees west of sub 1, although some are very widely dispersed. This position coincides with the brightest part of the nebula, known as the Hourglass Nebula. The massive star Herschel 36, which is the primary ionizing source of the nebula, is also a member of sub 3.
sub 4 is relatively distant from the other substructures, clustering at the edge of the Lagoon Nebula, about 0.4
degrees west of the cluster.
Meanwhile, the members of sub 5 are dispersed throughout the whole field and have no distinct center. They are located higher up in the binary tree, where the gravitational potential well is shallower, and they are the least bound members of the cluster. Thus, they are more likely to be contaminated by background stars.

\begin{figure}
\centering
\includegraphics[width=1\linewidth]{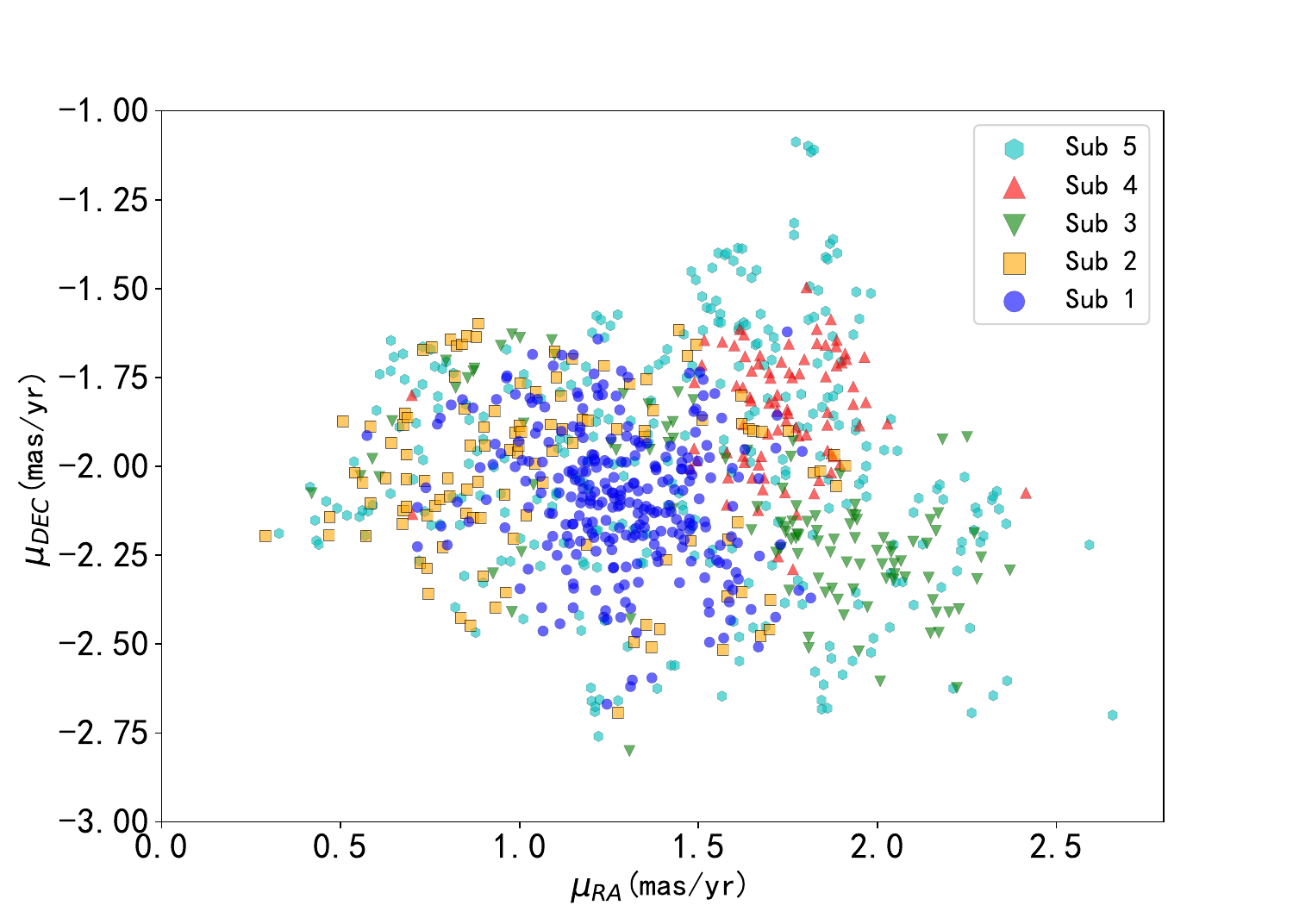}
\caption{\label{fig:pm}The proper-motion distributions of the five substructures. 
The symbols and color code are the same as Figure \ref{fig:space}.}
\end{figure}

Figure \ref{fig:pm} shows the proper-motion distribution of these five substructures.  
The situation is similar to their spatial distribution.
The members of sub 1 are concentrated in a relatively small area. 
Sub 2 members are located around sub 1 and are more dispersed. Since both sub 1 and sub 2 are located in the center of the field of view and form the main body of NGC 6530, we present some properties of these two structures combined in Table \ref{tab:subs} that are consistent with the results given by other studies. Both sub 3 and sub 4 are concentrated on the same side of sub 1, with a slightly larger $\mu_{R.A.}$. Conversely, the members of sub 5 are uniformly distributed among them.

\begin{figure}
\centering
\includegraphics[width=1\linewidth]{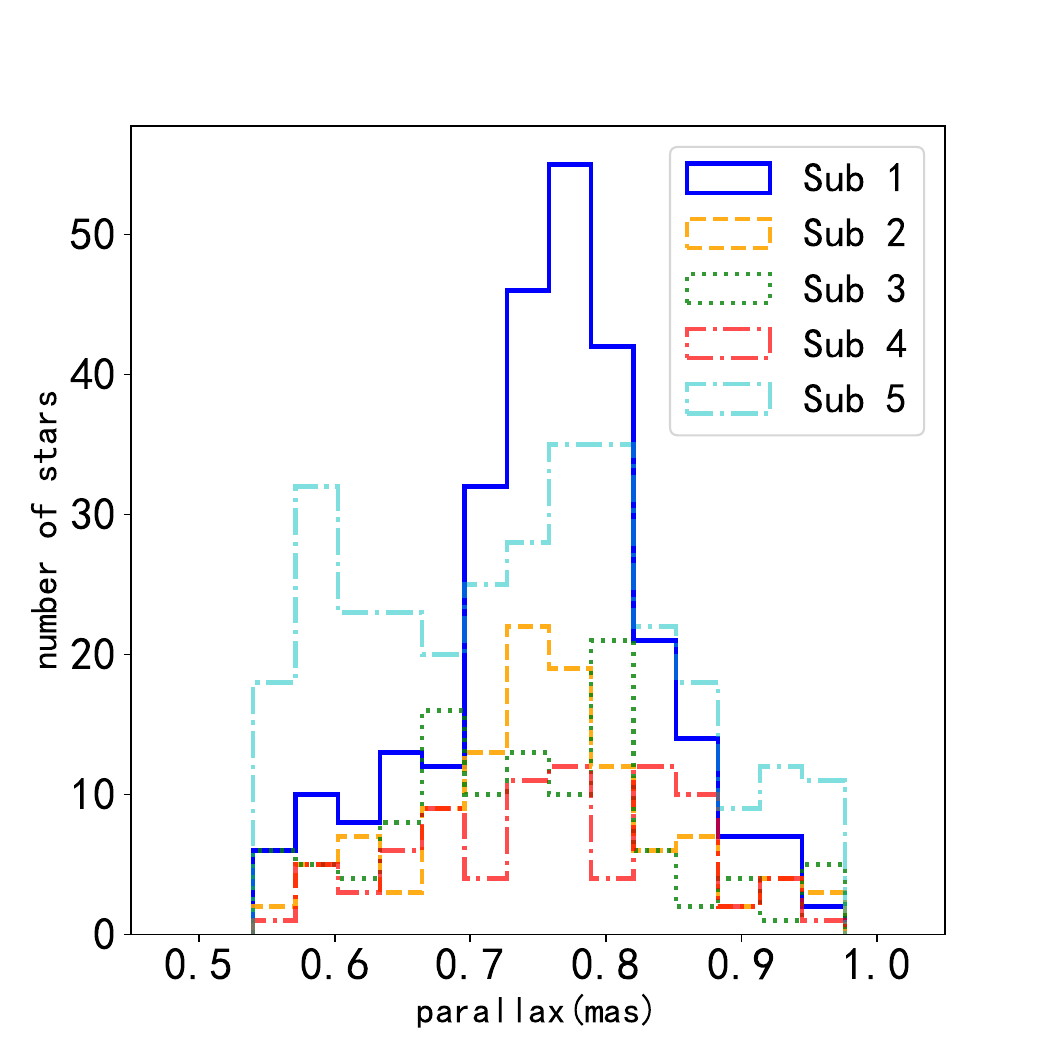}
\caption{\label{fig:para}The parallax distributions of the five substructures. Subs 1, 2, 3, and 4 all all roughly conform to a Gaussian distribution, but sub 5 presents a bimodal distribution. This suggests that it is contaminated by background structure.}
\end{figure}

To further investigate the distances of the different substructures, we check their parallax distributions as shown in Figure \ref{fig:para}.
For Sub 1, as core members of NGC 6530, its parallax is $0.757 \pm 0.083  $ mas, which is consistent with  the result of  \cite{2019MNRAS.486.2477W} of $0.724 \pm 0.186$ mas.  
Its standard error is not physical but mainly arises from the systematic uncertainty of the data. 
Regarding sub 2, sub 3, and sub 4, their members are less centralized but still conform to Gaussian distributions,
with mean values close to those of sub 1. However, sub 5 presents a bimodal distribution. It has a peak around 0.75 mas, accompanied by a second peak at approximately 0.6 mas, suggesting possible contamination by background stars.

In general, sub 1 contains the most stars in the central region and is compact in both the spatial distribution and the proper-motion space. It is the core of the cluster.
Sub 2 consists of less compact members in the core region. Both sub 3 and sub 4 are located west of the core. Their proper motions also differ  slightly from the core, suggesting different origins. 
Sub 5 contains members on the outskirts but is contaminated by background structures.

\section{Kinematic tracing}\label{sec:trace}

With the membership of substructures, we could also explore their kinematic properties.

\subsection{Local tracing}

The high-precision proper motion provided by Gaia DR3 allows us to trace the historical trajectories of every star over short periods.
Previous studies primarily utilized runaway stars or unbound stars for tracking purposes. \cite{2020ApJ...900...14F} and \cite{2022MNRAS.510.3178S} used runaway stars to investigate the historical information of the Orion Nebula Cluster. \cite{2023A&A...670A.128A}  studied simulation data and found that unbound stars contain more crucial early cluster information.
Additionally, for less-bound associations, \cite{2023MNRAS.522.3124Q} used all members of Auriga for tracing.
The substructures we obtained here using different thresholds correspond to members of the cluster with different degrees of binding. Their kinematic information can be used to trace the evolution of the cluster.

Considering open clusters disintegrate gradually after formation, especially in the early stages \citep{2023arXiv231202263D,2023A&A...673A.128G}, we can evaluate the kinematic age of the structures based on their size.
In this context, the half-mass radius $r_{50}$ is employed to represent the projected spatial size of the substructure. The physical meaning of the half-mass radius is the radius that contains half of the cluster members, providing an effective reflection of the cluster's size and density.
We adopt the median of all members' positions as the center of the structure.
Then we calculate the distances to the center of all members and use the median value as the $r_{50}$. 

\begin{figure}
\centering
\includegraphics[width=1\linewidth]{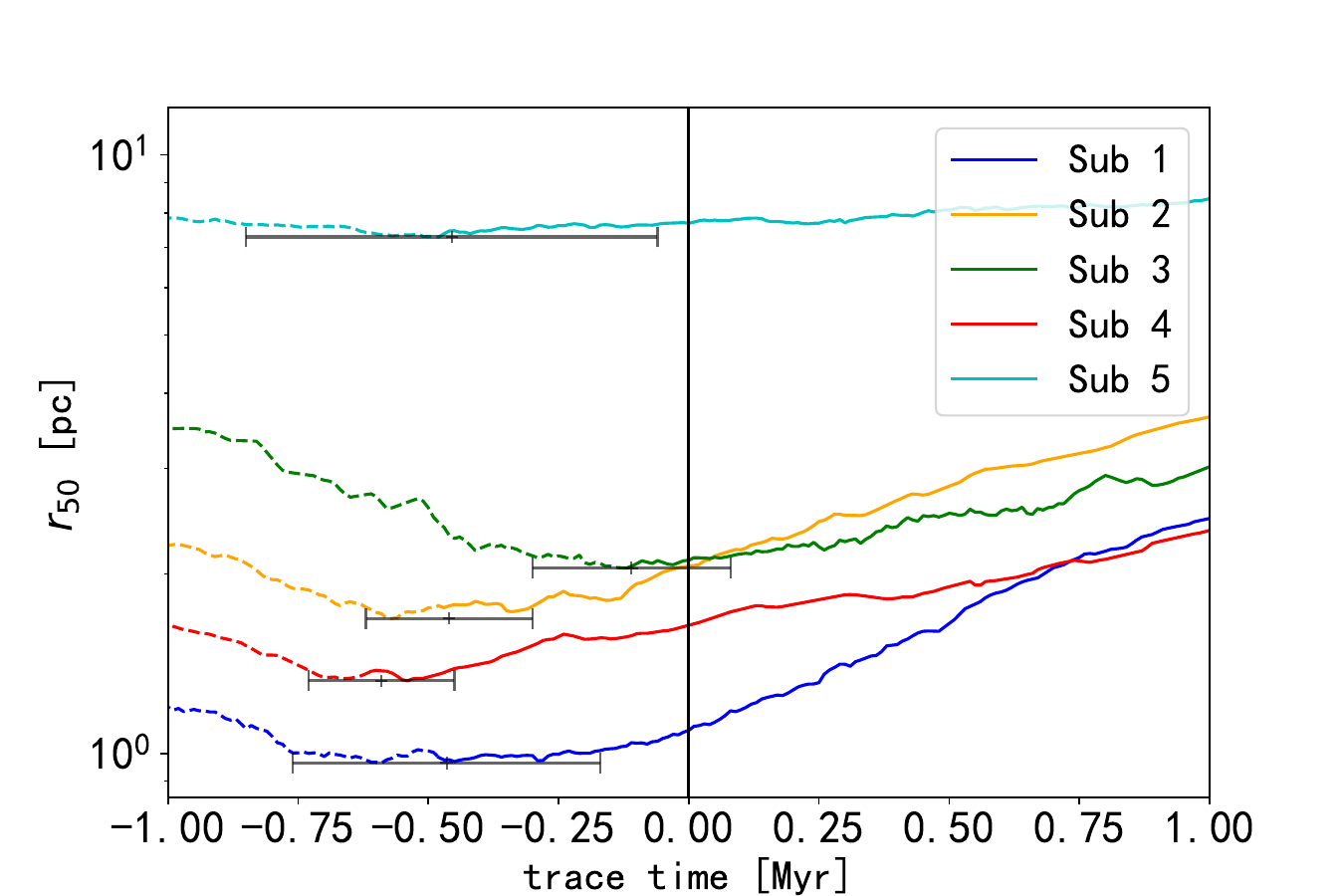}
\caption{\label{fig:trace}
Trends in the half-mass radius $r_{50}$ of the five substructures over time. The x-axis represents time, with positive values indicating the future and negative values indicating the past. The y-axis represents the spatial size. The cross is the median of the residence time. The dashed segments of the tracing lines indicate the trend in the absence of gravity.}
\end{figure}

With the positions and proper motions of the member stars, we trace each substructure to check the change of $r_{50}$ over a time scale. 
The results are shown in Figure \ref{fig:trace}.
However, we simply perform a linear tracing based on the current kinematic information of the member stars, without considering the effect of gravity. Their behavior after crossing the extremes is therefore unreliable and is plotted with dashed lines in Figure \ref{fig:trace}.

In order to characterize the time interval in which substructures remain at their minimum size, we define the time range in which the $r_{50}$ does not exceed 1.05 times the minimum $r_{50}$ as the residence time. This time range is more reasonable than the radius-minimum moment,
which is sensitive to the movement of specific members.
Table \ref{tab:info-4} provides a summary of the minimum $r_{50}$ values observed for each substructure, together with the corresponding residence time.

\begin{table}[htpb]
\centering
\caption{
The tracing results of \ref{fig:trace}.
The minimum half-mass radius $r_{50}$, and their corresponding residence time of the five substructures. }
\label{tab:info-4}
\centering
\begin{tabular}{|c|c|c|}
\hline
Sub ID & Min $r_{50}$ & Residence Time \\
  – &  (pc)    & (Myr)  \\
\hline
Sub 1      & 0.98     & $-0.46 \pm 0.30$\\ \hline
Sub 2      & 1.68     & $-0.46 \pm 0.16$\\ \hline
Sub 3      & 2.04     & $-0.11 \pm 0.19$\\ \hline
Sub 4      & 1.32     & $-0.59 \pm 0.13$ \\ \hline
Sub 5      & 7.30     & $-0.46 \pm 0.40$\\ \hline
\end{tabular}
\end{table}

According to the tracing results, the residence time of sub 1 is $-0.46 \pm 0.30 $ \rm{Myr}, while for sub 2, it's $-0.46 \pm 0.16 $ \rm{Myr}. These two substructures have the same kinematic ages, indicating a relatively similar time of origin. On the other hand, the size of sub 2 is always larger than that of sub 1. This implies that the  members of sub 2 were born in a larger region.

For sub 3, its residence time $-0.11 \pm 0.19 $ \rm{Myr} is close to zero
, indicating a relatively short kinematic age. Given that its spatial distribution overlaps with the young Hourglass Nebula, and that many members are located inside the star-forming region, it can be concluded that the stars of sub 3 originate from the Hourglass Nebula.
Sub 4 is located in a cavity-like region at the edge of the Lagoon Nebula. It has a residence time of $-0.59 \pm 0.13 $ Myr, which is further back in time than sub 3. 
The lack of gas around it indicates that it has passed the gas expulsion stage.
The residence time for sub 5 is $-0.46 \pm 0.40 $ Myr, a range that roughly includes the residence time for the other substructures.
However, due to the highly dispersed member distribution, the minimum $r_{50}$ of sub 5 is much larger than the $r_{50}$ values of other substructures.
Furthermore, there is no significant variation in this value.

It is important to note that the kinematic age estimation with linear tracing  is only applicable to young clusters. Typically, 
the cluster becomes virialized after a few freefall times 
\citep{2021MNRAS.506.5781L}. Consequently, the proper motions of the member stars are merely projected components of rotational velocity and could not be used for tracing. 
The freefall time $t_{ff}$ \citep{2012MNRAS.426.3008K} is defined as

\begin{equation}
    t_{ff}=\sqrt{\frac{{\pi}^2{r}^{3}_{\rm{tot},i}}{8G{M}_{\rm{tot},i}}}
\end{equation}

The variables ${M}_{\rm{tot},i}$ and ${r}_{\rm{tot},i}$ represent the initial total mass and radius of the star cluster.

We take the distance of the farthest star from the center at the moment of minimum sub1-2 radius as the initial radius (4.76 pc, 0.20°).
And, using the mass of the central part of the cluster of 3800 $M_{\odot}$ given using velocity dispersion \citep{2019MNRAS.486.2477W}, we could derive $t_{\rm{ff}}$= 2.79 Myr.
This means our kinematic linear tracing is valid at least for 1 or 2 million years before and after the present.

\subsection{Global tracing}\label{sec:Gtrace}
\begin{figure}
\centering
\includegraphics[width=1\linewidth]{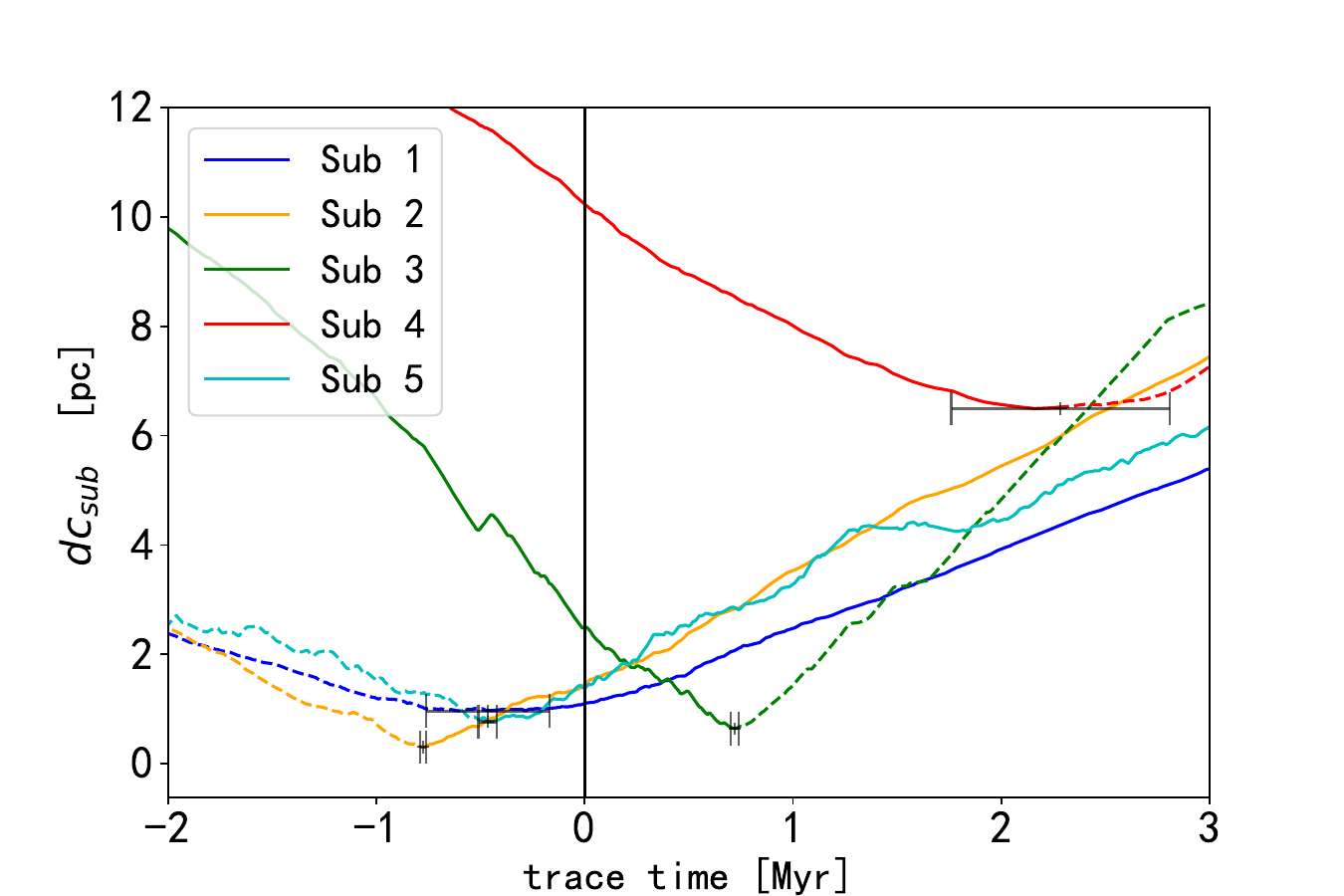}
\caption{\label{fig:ex}The trend of distances from each substructure to sub 1 ($dc_{sub}$). The blue line of sub 1 still represents the $r_{50}$ trend as a reference. The cross is the median of the minimum $dc_{sub}$ time range. The specific tracing results are given in Table \protect{\ref{tab:info-ex}}.  The dashed segments of the tracing lines indicate the trend in the absence of gravity.}
\end{figure}

In order to gain insight into the evolution of NGC 6530,
we also trace the distances between the substructures over time.
Given that sub 1 occupies a dominant position, we take its center as the reference point.
We denote the distance from the center of the substructure to the center of sub 1 as $dc_{sub}$.
The variation trend of $dc_{sub}$ over time is shown in Figure \ref{fig:ex}.
The $r_{50}$ trend of sub 1 is also provided as a reference (the blue line). 
As we do in the local tracing, we adopt 1.05 times minimum distance as a threshold to estimate the error of the distance-minimum moment.
Table \ref{tab:info-ex} presents the minimum distances, the corresponding distance-minimum moment, and the proportion of members of each substructure within Sub 1's $r_{90}$ (radius containing 90\% of members, about 2.22 pc at present) when their centers are closest.

\begin{table}[htpb]
    \centering
    \caption{
    The tracing results of \ref{fig:ex}.
    The table consists of four columns, including the substructure name, the minimum distance from that structure to Sub 1 (min $dc_{sub}$), the corresponding distance-minimum moment and its error, and the proportion of members of that structure within Sub 1's $r_{90}$ (radius containing 90\% of members) when their centers are closest. }
    \label{tab:info-ex}
    \begin{tabular}{|c|c|c|c|}
        \hline
        Sub ID & Min $dc_{Sub}$       &  Min $dc_{Sub}$ Moment &   $< r_{90}$ \\
        – &  (pc)    & (Myr)& (\%) \\
        \hline
        Sub 2      & 0.31    & $-0.78 \pm 0.01$ & 61\\ \hline
        Sub 3      & 0.64    & $0.72 \pm 0.02$ & 67\\ \hline
        Sub 4      & 6.50    & $2.28 \pm 0.52$ & 57\\ \hline
        Sub 5      & 0.77     & $-0.47 \pm 0.04$& 09\\ \hline
    \end{tabular}
\end{table}

The distance between the centers of sub 2 and sub 1 reached its closest point 780,000 yr ago ($-0.78 \pm 0.01 $ Myr).  This moment is not far from their residence times.
At that moment, the distance was even less than the $r_{50}$ of sub 1. Furthermore, 61\% of the members were located within the $r_{90}$ of sub 1.
This indicates that sub 2 and sub 1 share a similar origin.

The closest time between the centers of sub 3 and sub 1 is $0.72 \pm 0.02 $ Myr. 
At that moment, 65\% of the sub 3 members will be in $r_{90}$ of sub 1.
This indicates sub 3 is moving towards sub 1 and will fall into it in the future.
Sub 4 is also moving toward sub 1. However, our tracing results show it might not join sub 1. After 2.28 Myr, when the distance between their centers reaches its closest proximity at 6.50 pc, nearly half of its members will be within $r_{90}$ of sub 1.
If there is not enough gravity to bind them by then, they will not merge.

As for sub 5, its center had also been very close to the center of sub 1.
But very few of its members (only 9\%) were located within the $r_{90}$ radius of sub 1.
Thus, we conclude that its members are approximately uniformly distributed in space, with no apparent tendency to spread or accumulate.

The analysis results of \cite{2023A&A...670A.128A} on simulated data indicate that the use of unbound stars for tracing can provide more accurate information about the star cluster. 
However, the situation in NGC 6530 is not the case. 
Among the substructures we selected, sub 1 and sub 2 include most central members of the NGC 6530 cluster, whereas sub 5 is composed of less-bound members. Our tracing results show that sub 2 can better constrain the kinematic age than the less-bound sub 5. 
This is because NGC 6530 is a very young cluster. Its internal members have not evolved long enough to have lost early kinematic information. The escaping members have also not moved far enough away from the center. A proper motion of 1 $\rm{km~s^{-1}}$ causes only 0.28 degrees of positional shift in 1 million yr.
Perhaps in older clusters, unbound stars could provide more accurate kinematic information than members.
However, identifying former members in the outskirts of clusters is still challenging.

\section{CMD}\label{sec:diss}

Besides positions, proper motions, and distances, cluster members also share similar ages. 
Members of a cluster with different masses are distributed on a narrow main sequence of the CMD due to different evolutionary rates. 
This distribution could be used to constrain the ages of the substructures.

\begin{figure*}[htp]
\centering
\includegraphics[width=1\linewidth]{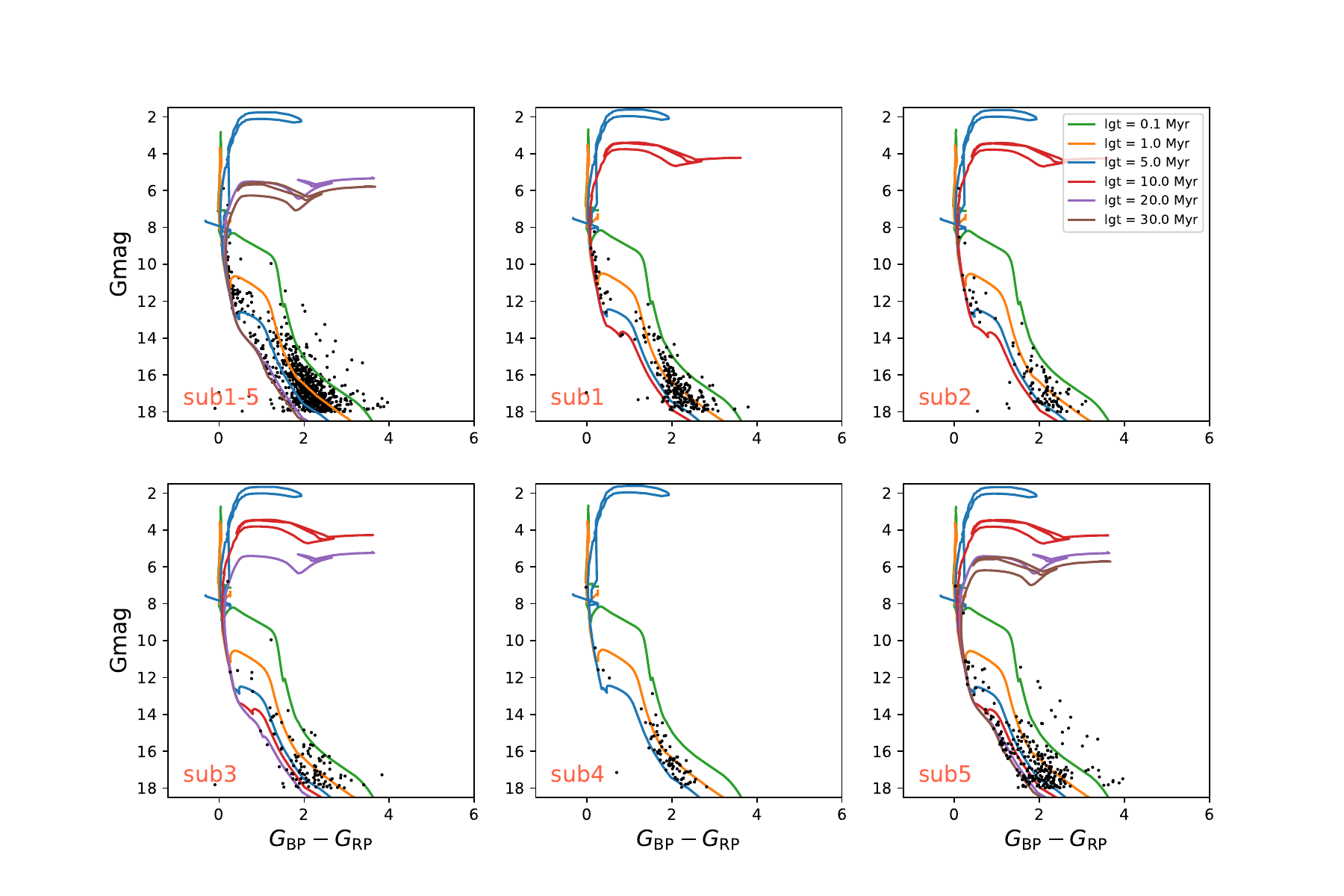}
\caption{\label{fig:hr}
Color-magnitude diagrams of the cluster (top left) and its five substructures. Isochrones of different ages are represented by different colors in the plot. }
\end{figure*}

We estimated the upper limits of the ages of the substructures by plotting isochrones of different ages on the CMDs of each substructure. We use the PARSEC theoretical isochrones version 1.2S \citep{2012MNRAS.427..127B} and the Gaia EDR3 photometric system \citep[][, $G$ and $G_{BP}$ - $G_{RP}$]{2021A&A...649A...3R}. In addition to age, we used the following parameters: the metallicity is fixed at solar metallicity, the distance modulus is derived using 
$m-M=5\log{d}-5=5\log{(1336\rm{pc}/10\rm{pc})}$ ; and $A_{V}$ = 1, estimated by eye-fitting the isochrones to the CMD. The CMDs of these substructures, as well as the isochrones for the different ages, are shown in Figure \ref{fig:hr} (15 stars without color information are not included).

The formation of a cluster is a process that takes place over tens of millions of years. For old clusters, this period is a small fraction of their total age, and all stars will have evolved to the main sequence. However, for young clusters (like this one), the formation process is a significant fraction of their age. Stars formed at different times will be at different stages of the pre-main-sequence. 
Additionally, the complex distribution of the interstellar medium in the young cluster (or star forming region) leads to different extinctions among cluster stars. 
These factors all cause a broadening of the main sequence on the CMD, making it difficult to determine a precise age for the cluster. Nevertheless, the left envelope of the main sequence can be used to estimate the upper limit of the age of the cluster stars, i.e., the earliest time at which star formation began in this (sub)cluster.

Numerous techniques have been developed for automatically fitting the isochrone in the CMD, e.g., BASE-9 \citep{2006ApJ...645.1436V}, 
ASteCA \citep{2015A&A...576A...6P}, fitCMD \citep{2019MNRAS.483.2758B},
MiMO \citep{2022ApJ...930...44L},  and Bayesian convolutional neural network  \citep{2023A&A...673A.114H}. These software packages are designed for star clusters with a single age and uniform external extinction, otherwise the derived ages would be inaccurate. Therefore, in this work, we use the eye-fitting the left envelope to estimate the earliest star formation time of the cluster (or subcluster). As shown in Figure \ref{fig:hr}, the estimated upper ages of these substructures are sub 1: 10 Myr; sub 2: 6-7 Myr; sub 3 and sub 4: 20 Myr; sub 5: 30 Myr. Considering that this is only an eye-fit, we believe that a relative error of $10\%$ is reasonable for this upper limit.

The upper ages derived from the isochrones are generally older than the kinematic ages obtained with linear tracing. This suggests that the kinematic tracing approach is not very effective for determining the age of the oldest stars in a cluster. 
The contraction and expansion of the substructures they contain may be repeated many times during the evolution of the cluster.
Thus, these ages cannot be taken as the time of origin of these substructures. Nevertheless, the recent structural trends revealed by this method are still crucial to our understanding of the evolution of clusters. Considering that there are many stars in each substructure distributed in isochrones between 0.1 Myr and 1 Myr, there is no conflict between our kinematic ages and the age distribution of the stars.

\section{Conclusions}\label{sec:conclusion}

The stars in open clusters are not born in the same place or at the same time. Clusters also gradually fall apart under the influence of their environment. To explore this complex process, we employ the hierarchical clustering algorithm to identify the internal structures of the young open cluster NGC 6530.

Our algorithm groups members of the cluster into 5 substructures: sub 1 as the core of the NGC 6530 star cluster, sub 2 is similar to sub 1, but spread over a larger region. sub 3 is associated with a young star forming region -- the Hourglass Nebula; sub 4 is an outskirt group; while sub 5 is composed of less-bound members with field contamination.

We also explore the evolutionary trends of these substructures with linear kinematic tracing of their members.
By evaluating the sizes ($r_{50}$) of the substructures over time (local tracing), we find that the sizes of sub 1 , sub 2, and sub 4 all reach their minima around 0.5 Myr, which is consistent with the age estimate of the cluster.
However, sub 3 reaches its minimum size around 0.1 Myr.  This indicates that it is a young component of the cluster.

By examining the trends in distances between the centers of these substructures (global tracing), we explore the evolutionary process of the substructures.
The kinematic time of sub 2 could be traced back to 0.78 Myr ago.
At that time, 61\% of its members were located within $r_{90}$ of Sub 1. They gradually migrated outwards later.
The newly formed Sub 3 will merge with the core after 0.72 Myr. Sub 4 is also moving towards the core, but might not end up merging.  This suggests that it may have originated independently in the outskirt region of the molecular cloud.

We also estimated the upper limits of the ages of the substructures by plotting isochrones of different ages on the color-magnitude diagrams. Due to the complexity of the extinction of the molecular cloud, the age obtained by this method is very uncertain. 
However, the kinematic ages are still consistent with the age distribution of the substructure members.

With the hierarchical clustering algorithm, we not only reveal the complex internal structure of the young cluster NGC 6530, but also provide a clear picture of the internal relationships of this cluster, presenting a concrete description of its past and future evolution. This method can help us to better analyze and understand the evolution of star clusters.

\begin{acknowledgments}
We would like to thank the anonymous referee for valuable comments and suggestions that greatly improved the quality of this work. Zhengyi Shao acknowledges the National Natural Science Foundation of China (NSFC) under grants 12273091, U2031139, and the partly supports of the National Key R\&D Program of China No. 2019YFA0405501,  the science research grants from the China Manned Space Project with No. CMS-CSST-2021-A08 and the Science and Technology Commission of Shanghai Municipality (Grant No. 22dz1202400).
Li, L. thanks the support of NSFC No. 12303026 and the Young Data Scientist Project of the National Astronomical Data Center.

This work has made use of data from the European Space Agency (ESA) mission Gaia ( \href{https://www. cosmos.esa.int/gaia}{https://www. cosmos.esa.int/gaia} ), processed by the Gaia Data Processing and Analysis Consortium (DPAC, \href{https://www.cosmos.esa. int/web/gaia/dpac/consortium}{https://www.cosmos.esa. int/web/gaia/dpac/consortium} ). Funding for the DPAC has been provided by national institutions, in particular the institutions participating in the Gaia Multilateral Agreement. This research has made use of NASA’s Astrophysics Data System Bibliographic Services. The optical images used in this study are from the Digitized Sky Survey II (DSS2, \href{https://archive.eso.org/dss/dss}{https://archive.eso.org/dss/dss} ). 

\software{astropy \citep{2013A&A...558A..33A,2018AJ....156..123A}, Scikit-learn\citep{2011JMLR...12.2825P}}

\end{acknowledgments}

\appendix
\section{Background structure}\label{sec:appendixa}
\renewcommand{\thefigure}{A.\arabic{figure}}

We apply the hierarchical clustering algorithm to  the initial dataset, which contains 20,819 stars. According to the weight values ($w_0$) of the $\sigma$-plateau with different $p$, we find that the optimized value of the parameter $p$ is 10. 

\begin{figure}[htp]
\centering
\includegraphics[width=0.9\linewidth]{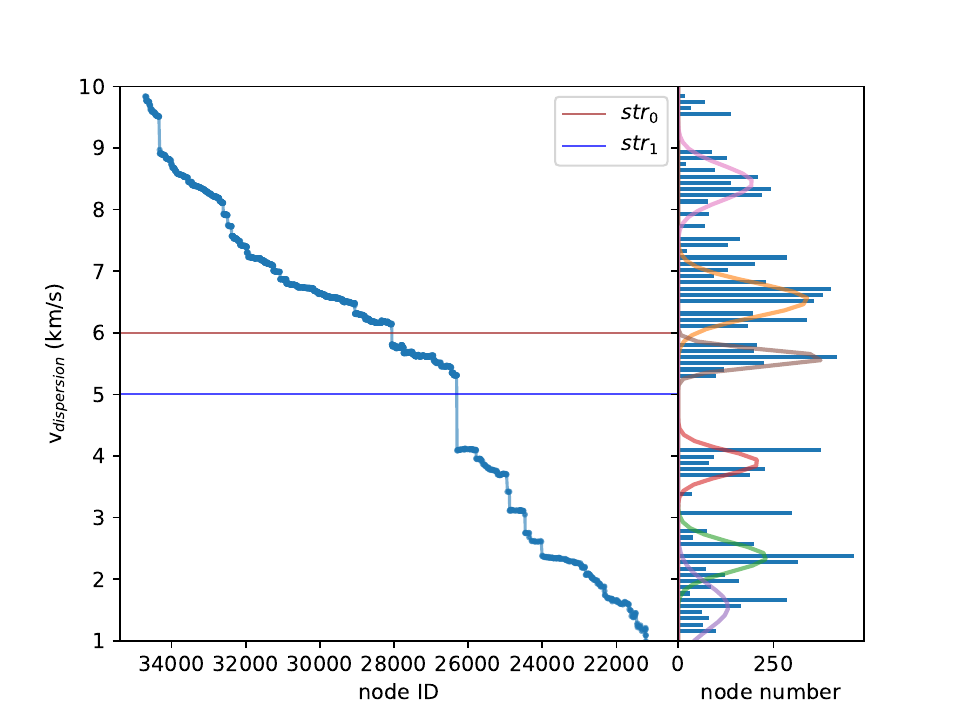}
\caption{\label{fig:vhst-all}
Similar to Figure \ref{fig:vhst}, but the dataset is 20,819 stars. The two lines from bottom to top correspond to 5 $\rm{km~s^{-1}}$ and 6 $\rm{km~s^{-1}}$ respectively.}
\end{figure}

Then we identify two structures in the field, called $str_{0}$ and $str_{1}$. The corresponding velocity dispersion profile is shown in Figure \ref{fig:vhst-all}. The two lines from bottom to top correspond to 5 $\rm{km~s^{-1}}$ and 6 $\rm{km~s^{-1}}$ respectively. The structure between 6 and 5 is $str_{0}$, below 5 is $str_{1}$.

\begin{figure}[htp]
\centering
\includegraphics[width=0.9\linewidth]{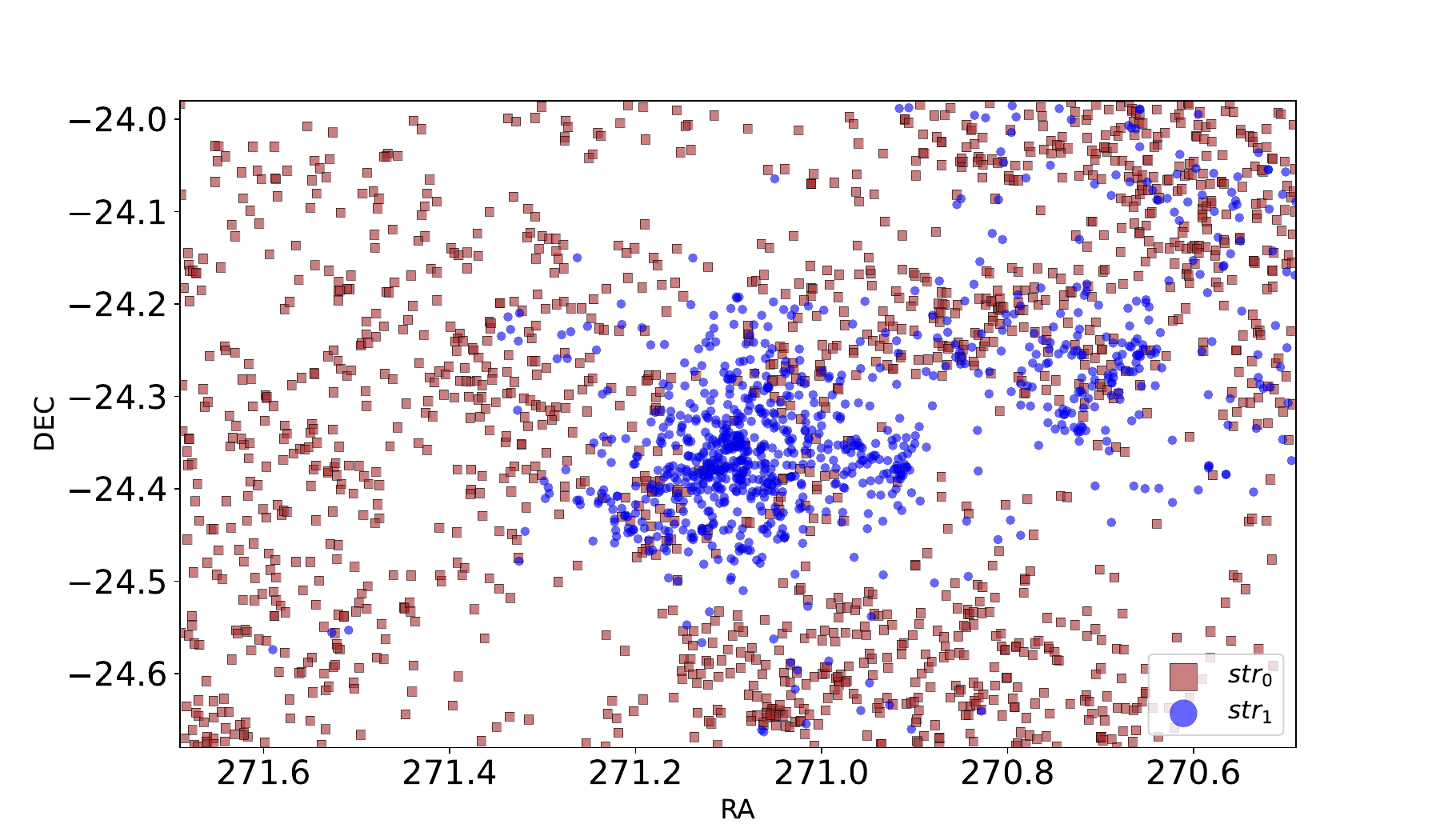}
\caption{\label{fig:space-all}
The spatial distribution of the two structures, with blue dots representing $str_{1}$ and brown squares indicating $str_{0}$.}
\end{figure}

\begin{figure}[htp]
\centering
\includegraphics[width=0.9\linewidth]{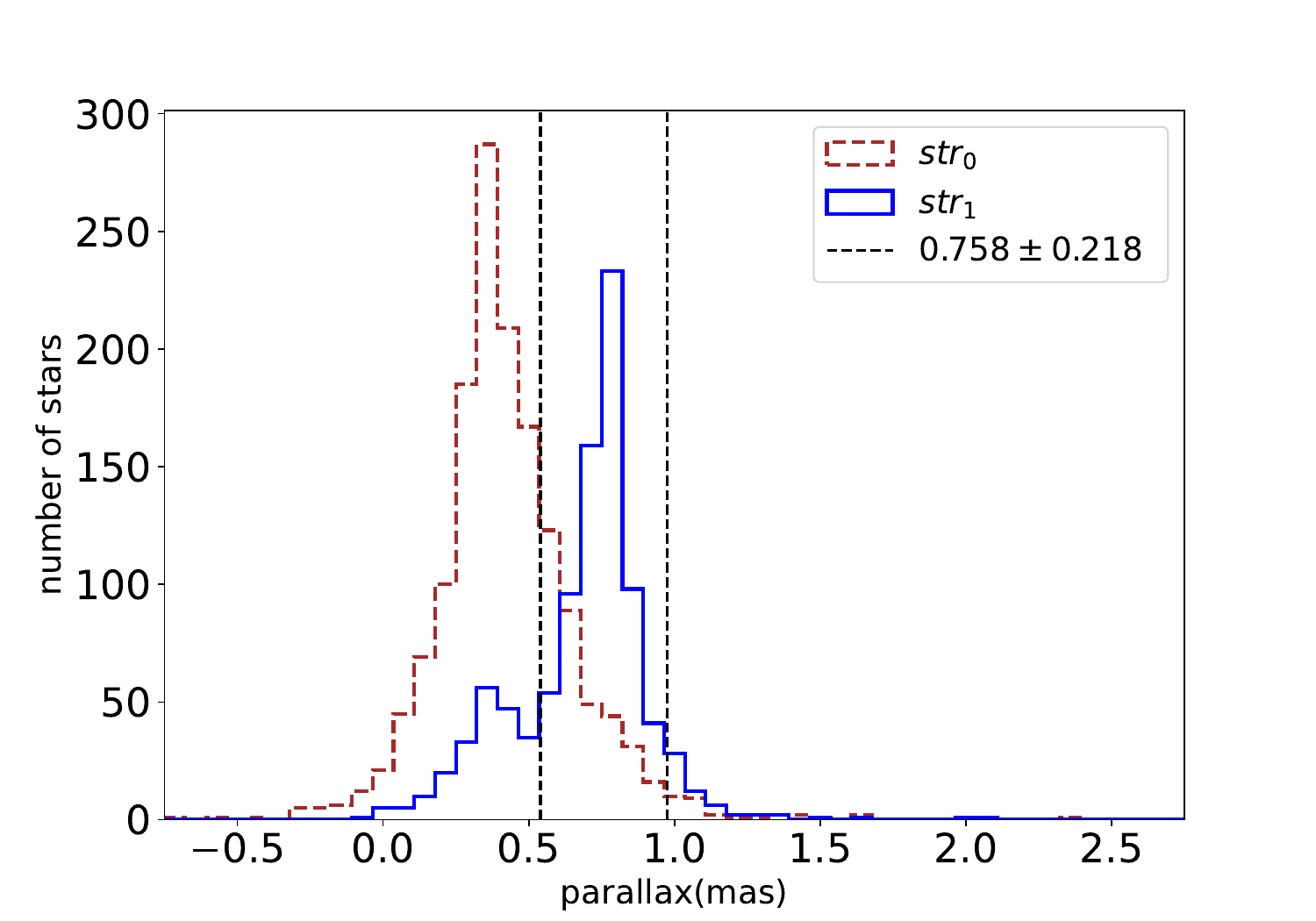}
\caption{\label{fig:para-all}
Parallax distribution of the two structures. The dashed brown line represents $str_{0}$. While the solid blue line represents $str_{1}$. The black vertical dashed lines represent the range of parallax constraints.}
\end{figure}

The spatial distributions of these two structures are shown in Figure \ref{fig:space-all}.
$str_{0}$ is the main structure in the field, but appears as a diffuse distribution throughout the entire field of view, while $str_{1}$ is located in the center of the field, correlating with NGC 6530. The parallax distribution of their members is shown in Figure \ref{fig:para-all}. These two structures are located at different distances. $str_{0}$ has a parallax of 0.4 $\rm{mas}$ (2500 $\rm{pc}$), corresponding to the Scutum-Centaurus Arm of the Milky Way \citep{2021MNRAS.500.3064S}, while the parallax of $str_{1}$ is about 0.758 mas ( $\sim$ 1319 pc), consistent with the distance of NGC 6530.

It is worth noting that a small fraction of the members of $str_{1}$ are actually background stars at the distance of $str_0$.
Considering the large number of stars at the distance of $str_0$ and spread over the entire field of view, the presence of these background stars can significantly affect the subsequent analysis. We then further filtered the dataset according to the parallax distribution of the $str_1$ members: $0.758 \pm 0.218 $ \rm{mas}.
In the end, we obtain 5,945 stars for further analysis.

\bibliography{sample631}{}
\bibliographystyle{aasjournal}

\end{CJK*}
\end{document}